\documentclass[12pt]{article}

\usepackage[dvipdfmx]{graphicx}
\usepackage[dvipdfmx]{xcolor}
\usepackage{amsmath}
\usepackage{amssymb}
\usepackage{tikz}
\usepackage{cite}
\usepackage{fancyhdr}

\newcommand{\Slash}[1]{{\ooalign{\hfil/\hfil\crcr$#1$}}}

\def\erase#1{{}}
\def\EqArrerase#1{{}}

\makeatletter
 
 \@addtoreset{equation}{section}
\makeatother

\def\GL{{G\kern-.12em L\kern-.04em}}
\def\OSp{{O\kern-.11em S\kern-.04em p}}
\def\IOSp{{I\kern-.06em O\kern-.11em S\kern-.04em p}}
\def\MN{{M\kern-.14em N}}
\def\NM{{N\kern-.14em M}}
\def\NL{{N\kern-.14em L}}
\def\LN{{L\kern-.11em N}}
\def\ML{{M\kern-.14em L}}
\def\LM{{L\kern-.11em M}}
\def\RN{{R\kern-.11em N}}
\def\NR{{N\kern-.14em R}}
\def\RM{{R\kern-.11em M}}
\def\MR{{M\kern-.14em R}}
\def\RL{{R\kern-.11em L}}
\def\LR{{L\kern-.11em R}}
\def\RS{{R\kern-.11em S}}
\def\SR{{S\kern-.11em R}}
\def\SN{{S\kern-.11em N}}
\def\NS{{N\kern-.11em S}}
\def\SM{{S\kern-.11em M}}
\def\MS{{M\kern-.11em S}}
\def\SL{{S\kern-.11em L}}
\def\LS{{L\kern-.11em S}}
\def\sqr#1#2{{\vcenter{\hrule height.#2pt
      \hbox{\vrule width.#2pt height#1pt \kern#1pt
          \vrule width.#2pt}
      \hrule height.#2pt}}}
\def\bra0{\langle0|}
\def\ket0{|0\rangle}
\def\soeji#1_#2#3{#1_{#2}\cdots#1_{#3}}
\def\longgLRarrow{\longleftarrow\kern-3pt\relbar\kern-3pt\relbar\kern-3pt%
\longrightarrow}
\def\longLRarrow{\longleftarrow\kern-3pt\relbar\kern-3pt\longrightarrow}
\def\longLarrow{\longleftarrow\kern-3pt\relbar\kern-3pt\relbar\kern-3pt\relbar}
\def\longRarrow{\relbar\kern-3pt\relbar\kern-3pt\relbar\kern-3pt\longrightarrow}
\def\bothDer#1#2#3{%
\overset{\kern-.7em\stackrel{#1}{#2}}{\partial_{#3}}}
 
\makeatletter
 
 \@addtoreset{equation}{section}
\makeatother

\begin{document}
\thispagestyle{fancy}

\title{Color Confinement and Massive Gluon in Superfield Formalism}

\author{Ichiro Oda
\footnote{Electronic address: ioda@cs.u-ryukyu.ac.jp}
\\
{\it\small
\begin{tabular}{c}
Department of Physics, Faculty of Science, University of the 
           Ryukyus,\\
           Nishihara, Okinawa 903-0213, Japan\\      
\end{tabular}
}
}
\date{}

\maketitle

\thispagestyle{fancy}

\begin{abstract}

In connection with the question of confinement of massive ghost in quadratic gravity (QG),
color confinement in quantum chromodynamics (QCD) has been reconsidered  
in the superfield formalism. It is shown that when a bound state in the BRST transformation 
of the gluon field exists, the gluon becomes massive and is confined. It is also shown that 
the asymptotic field of the gluon field obeys the field equation for not the conventional massive 
Klein-Gordon field but the massive dipole field. In case of quark confinement, it is shown that 
the quark field satisfies the massive spinor dipole equation in the confinement phase, 
which might suggest a physical picture such that a pair of quark and anti-quark constitutes 
a bound state and is confined into a meson. These facts encourage us to conjecture that 
the similar phenomenon could take place in confinement of massive ghost, which violates 
the unitarity of the physical S-matrix, and might provide us with a resolution to the unitarity 
problem in QG.   

\end{abstract}

\newpage
\pagestyle{plain}
\pagenumbering{arabic}


\section{Introduction}

It is well-known that quantum chromodynamics (QCD) and quadratic gravity (QG) possess common physical features.
For instance, they are asymptotically free\footnote{See \cite{GW, Pol} for QCD and \cite{Mario, Fradkin, Avramidi1, Avramidi2, Savio} for QG.} 
and renormalizable\footnote{See \cite{Hooft1, Hooft2} for QCD and \cite{Stelle} for QG.} 
so they can become candidates of the ultraviolet(UV)-complete theory. 
In QCD, one of the central problems is to prove color confinement. 
Because of the similarity between QCD and QG, it is natural to conjecture that there might be a confinement phase in QG as well 
where massive ghost, which violates the unitarity of the physical S-matrix, could be confined by the gravitational interaction. 

Indeed, in our recent work \cite{Oda-MG}, based on the superfield formalism by Bonora and Tonin \cite{Bonora-Tonin} 
we have shown that if there is a tensor bound state in the BRST transformation of the massive ghost field, 
together with the other members of the same BRST quartet, the asymptotic state of the massive ghost appears 
in the physical subspace only in the zero-norm combinations and in essence decouples from physical Hilbert space \cite{Kimura}. 
This mechanism of confinement of the massive ghost could provide us with a resolution to the infamous problem
of unitarity violation in QG. Furthermore, we have demonstrated that even if the asymptotic state of the massive ghost obeys 
the massless Klein-Gordon equation in the deconfinement phase, it becomes to satisfy the massive dipole equation 
in the confinement phase. 
 
In this article, we will exhibit that the similar phenomenon takes place even in QCD. 
Precisely speaking, the more interesting phenomena happen in QCD compared to QG:
In confinement of the gluon, the asymptotic field of the gluon becomes massive and obeys the massive
dipole equation. In other words, the gluon, which is originally massless, acquires a mass (or
mass gap emerges) at the Lagrangian level without recourse to some dynamical mechanism when color confinement
occurs. In case of quark confinement, the quark satisfies the massive Dirac dipole equation. The massive dipole field 
is equivalent to two massive Dirac simple pole fields, so this fact might imply a picture such that a pair of massive quark 
and anti-quark constitutes a bound state, what we call, ``meson.''    

The paper is organized as follows: In the next section, we present a brief review of a superfield formalism 
in the six-dimensional superspace by Bonora and Tonin \cite{Bonora-Tonin}. In Section 3, we explain 
confinement of color in QCD. In Section 4, we apply the superfield formalism to confinement of the gluon and quark
and derive effective Lagrangians and the field equations for the asymptotic fields. 
In final section, we draw our conclusion.

\section{Review of superfield formalism}

Prior to a review of the superfield formalism, let us recaptuate the BRST formalism of the Yang-Mills theory.
The standard Yang-Mills Lagrangian, which is supplemented by the covariant gauge fixing term and the Faddeev-Popov (FP)
ghost term, is given by
\begin{eqnarray}
{\cal{L}} = {\rm{Tr}} \Big( - \frac{1}{4} F_{\mu\nu} F^{\mu\nu} - B \partial^\mu A_\mu   
+ \frac{\alpha}{2} B^2 - i \partial^\mu \bar C D_\mu C \Big),
\label{YM-Lag1}  
\end{eqnarray}
where 
\begin{eqnarray}
F_{\mu\nu}^a &=& \partial_\mu A_\nu^a - \partial_\nu A_\mu^a + g f_{abc} A_\mu^b A_\nu^c,
\nonumber\\
D_\mu C^a &=& \partial_\mu C^a + g f_{abc} A_\mu^b C^c. 
\label{F&DC}  
\end{eqnarray}
Here $B, C$ and $\bar C$ are the auxiliary field, the FP ghost and the FP antighost, respectively, 
$\alpha$ is the gauge parameter, and $g$ is the coupling constant. Moreover, we use the notation, 
for instance, $A_\mu = A_\mu^a T^a$ where $T^a$ are generators of the Lie group 
which obey the commutation relations with the structure constant $f_{abc}$, i.e., $[ T_a, T_b ] = i f_{abc} T_c$. 
We also use the notation, $(A \times B)^a \equiv [A, B]^a \equiv f^{abc} A^b B^c$ in this article.

The Lagrangian (\ref{YM-Lag1}) is invariant under the BRST transformation $\delta_B$ generated 
by the BRST charge $Q_B$ \cite{Kugo-Ojima} and the anti-BRST transformation $\bar \delta_B$ 
generated by the anti-BRST charge $\bar Q_B$ \cite{Curci, Ojima}, which concretely take the form:
\begin{eqnarray}
\delta_B A_\mu^a &=& D_\mu C^a, \qquad
\delta_B C^a = - \frac{g}{2} ( C \times C )^a,
\nonumber\\
\delta_B \bar C^a &=& i B^a, \qquad
\delta_B B^a = 0, 
\label{BRST-transf}  
\end{eqnarray}
and
\begin{eqnarray}
\bar \delta_B A_\mu^a &=& D_\mu \bar C^a, \qquad
\bar \delta_B C^a = i \bar B^a,
\nonumber\\
\bar \delta_B \bar C^a &=& - \frac{g}{2} ( \bar C \times \bar C )^a, \qquad
\bar  \delta_B \bar B^a = 0. 
\label{anti-BRST-transf}  
\end{eqnarray}
Here the auxiliary fields $B$ and $\bar B$ are related by
\begin{eqnarray}
B^a + \bar B^a = i g ( C \times \bar C)^a,
\label{B-bar-B}  
\end{eqnarray}
from which we have the remaining (anti-)BRST transformation:
\begin{eqnarray}
\delta_B \bar B^a = g  ( \bar B \times C )^a, \qquad
\bar \delta_B B^a = g  ( B \times \bar C)^a.
\label{Re-BRST}  
\end{eqnarray}
The charges $Q_B$ and $\bar Q_B$ are nilpotent and anti-commuting with each other. 

Now we will briefly review the superfield formalism in the six-dimensional superspace 
parametrized by coordinates $(x^\mu, \theta, \bar \theta)$ where $\theta$ and $\bar \theta$ 
are two Grassmann numbers obeying $\theta^* = - \theta, \bar \theta^* = - \bar \theta, \theta^2 
= \bar \theta^2 = \{ \theta, \bar \theta \} = 0$ \cite{Bonora-Tonin}. Because of Grassmann nature 
of $\theta$ and $\bar \theta$, the superfield can be expanded into a finite Taylor series as
\begin{eqnarray}
\Phi ( x, \theta, \bar \theta ) = \left. \Phi \right|_0 
+ \theta \left. \frac{\partial \Phi}{\partial \theta} \right|_0 
+ \bar \theta \left. \frac{\partial \Phi}{\partial \bar \theta} \right|_0 
+ \bar \theta \theta \left. \frac{\partial^2 \Phi}{\partial \theta \partial \bar \theta} \right|_0,
\label{Superfield}  
\end{eqnarray}
where $\left. {} \right|_0$ denotes setting $\theta = \bar \theta = 0$. Provided that we can identify 
$\frac{\partial}{\partial \theta}$ and $\frac{\partial}{\partial \bar \theta}$ as the BRST transformation 
$\delta_B$ and the anti-BRST transformation $\bar \delta_B$, respectively, 
Eq. (\ref{Superfield}) can be rewritten as 
\begin{eqnarray}
\Phi ( x, \theta, \bar \theta ) = \Phi (x) + \theta \delta_B \Phi (x) + \bar \theta \bar \delta_B \Phi (x) 
+ \bar \theta \theta \delta_B \bar \delta_B \Phi (x).
\label{Superfield-2}  
\end{eqnarray}

The superfield formalism by Bonora and Tonin \cite{Bonora-Tonin} aims at this identification 
in a geometrical manner in the six-dimensional superspace. Namely, starting with the six-dimensional 
geometrical objects, we try to derive Eq. (\ref{Superfield-2}) by imposing some geometrical constraint 
which is sometimes called ``horizontality condition'', meaning the flatness condition of the field strengh
in the Grassmann directions \cite{Kugo-text}.\footnote{This condition is called ``soul flatness condition'' in \cite{N-O-text}.}

In the $(4 + 2)$-dimensional superspace parametrized by $z^M = (x^\mu, \theta, \bar \theta)$, 
one-form gauge field ${\cal{A}}(z)$ is defined as
\begin{eqnarray}
{\cal{A}}(z) \equiv {\cal{A}}_M (z) d z^M \equiv A_\mu (z) d x^\mu + A_\theta (z) d \theta + A_{\bar \theta} (z) d \bar \theta.
\label{Gauge-field}  
\end{eqnarray}
Then, the corresponding two-form field strength is defined by
\begin{eqnarray}
{\cal{F}}(z) \equiv \tilde d {\cal{A}} (z) - i g {\cal{A}}^2 \equiv - \frac{1}{2} F_{NM} (z) d z^M d z^N,
\label{Field-str}  
\end{eqnarray}
where one-form derivative $\tilde d$ is defined by
\begin{eqnarray}
\tilde d \equiv d + \delta + \bar \delta \equiv \frac{\partial}{\partial x^\mu} d x^\mu + \frac{\partial}{\partial \theta} d \theta
+ \frac{\partial}{\partial \bar \theta} d \bar \theta.
\label{Deriva}  
\end{eqnarray}
In calculating various quantities, we use the following relations with respect to the grading:
\begin{eqnarray}
d x^M d x^N &=& (-1)^{|M| |N| +1} d x^N d x^M, 
\nonumber\\
( x^M, \partial_M ) d x^N &=& (-1)^{|M| |N|} d x^N ( x^M, \partial_M ),
\label{Grading}  
\end{eqnarray}
where $|M|$ indicates that $|M| = 1$ when $x^M = \theta, \bar \theta$ while $|M| = 0$ when $x^M = x^\mu$. 

The horizontality condition, which is needed to make Eq. (\ref{Superfield}) coincide with Eq. (\ref{Superfield-2}),
is provided with 
\begin{eqnarray}
{\cal{F}}(z) = \frac{1}{2} F_{\mu\nu} (z) d x^\mu d x^\nu,
\label{Horizon}  
\end{eqnarray}
or equivalently,
\begin{eqnarray}
F_{M \theta} (z) = F_{M \bar \theta} (z) = 0.
\label{Horizon2}  
\end{eqnarray}

With ``boundary conditions'',
\begin{eqnarray}
\left. A_\mu (z) \right|_0 &=& A_\mu (x), \qquad
\left. A_\theta (z) \right|_0 = C (x), \qquad
\left. A_\theta (z) \right|_0 = C (x),
\nonumber\\
\left. \partial_\theta A_{\bar \theta} (z) \right|_0 &=& i B(x), \qquad
\left. \partial_{\bar \theta} A_\theta (z) \right|_0 = i \bar B(x),
\label{BC}  
\end{eqnarray}
the horizontality condition determines the gauge superfield to 
\begin{eqnarray}
A_\mu (z) &=& A_\mu (x) + \theta D_\mu C(x) + \bar \theta D_\mu \bar C(x) 
+ \bar \theta \theta ( i D_\mu B(x) + g D_\mu C(x) \times \bar C(x) ),
\nonumber\\
A_\theta (z) &=& C (x) + \theta ( - \frac{g}{2} C(x) \times C(x) ) + \bar \theta i \bar B(x) 
+ \bar \theta \theta i g \bar B(x) \times C(x),
\nonumber\\
A_{\bar \theta} (z) &=& \bar C (x) + \theta i B(x) + \bar \theta ( - \frac{g}{2} \bar C(x) \times \bar C(x) )
+ \bar \theta \theta (- i g ) B(x) \times \bar C(x).
\label{G-superfield}  
\end{eqnarray}
Moreover, the field strength superfield reads
\begin{eqnarray}
F_{\mu\nu} (z) &=& F_{\mu\nu} (x) + \theta F_{\mu\nu}(x) \times C(x) 
+ \bar \theta F_{\mu\nu}(x) \times \bar C(x) 
\nonumber\\
&+& \bar \theta \theta [ i F_{\mu\nu}(x) \times B(x) + ( F_{\mu\nu}(x) \times C(x) )
\times \bar C(x) ].
\label{F-superfield}  
\end{eqnarray}

Then, it is straightforward to rewrite the Yang-Mills Lagrangian (\ref{YM-Lag1}) in terms of 
the superfields as
\begin{eqnarray}
{\cal{L}} = - \frac{1}{4} F^a_{\mu\nu}(z) F^{a\mu\nu}(z) 
+ \frac{i}{2} \frac{\partial^2}{\partial \theta \partial \bar \theta} ( A^a_\mu(z) A^{a\mu}(z) )
- \frac{\alpha}{2} \left(\frac{\partial A^a_{\bar \theta}(z)}{\partial \theta} \right)^2.
\label{YM-Lag2}  
\end{eqnarray}
Here let us note that there is a nontrivial relation, $F^a_{\mu\nu}(z) F^{a\mu\nu}(z) 
= F^a_{\mu\nu}(x) F^{a\mu\nu}(x)$. Moreover, note that the last term on the RHS
does not have the form of $\frac{\partial^2}{\partial \theta \partial \bar \theta} ( \dots )$.
Instead of the last term, it is possible to introduce such a term like
\begin{eqnarray}
\frac{\partial^2}{\partial \theta \partial \bar \theta} ( A^a_\theta(z) A^a_{\bar \theta}(z) )
= \frac{1}{2} ( B^2 + \bar B^2 ).
\label{Rewrite}  
\end{eqnarray}
The meaning of a term involving $\bar B^2$ can be found in \cite{Bonora-Tonin}.

As for the spinor superfield $\Psi(z)$, the horizontality condition is of form:
\begin{eqnarray}
D_M \Psi(z) d z^M = D_\mu \Psi(z) d z^\mu,
\label{M-Horizon}  
\end{eqnarray}
or equivalently,
\begin{eqnarray}
D_\theta \Psi(z) = D_{\bar \theta} \Psi(z) = 0.
\label{M-Horizon2}  
\end{eqnarray}
This horizontality condition uniquely fixes the expression of $\Psi(z)$ as
\begin{eqnarray}
\Psi(z) &=& \Psi(x) + \theta ( i g C(x) \Psi(x) ) + \bar \theta ( i g \bar C(x) \Psi(x) )
\nonumber\\
&-& \bar \theta \theta g ( B(x) - g \bar C(x) C(x) ) \Psi(x).
\label{M-superfield}  
\end{eqnarray}

One of advantages in the superfield formalism by Bonora and Tonin is that the component field associated with 
$\bar \theta \theta$ automatically becomes physical observables since it is invariant under both BRST and 
anti-BRST transformations. On the other hand, one of unsatisfactory features is that we cannot introduce
the horizontality condition in the Lagrangian and must impose it by hand.

\section{Color confinement}

In this section, we shall present a mechanism for confinement of gluons and quarks. 
The key idea in this mechanism is that the concept of color confinement is closely related to 
the formation of bound states between the elementary fields such as gluons and quarks, 
and the Faddeev-Popov (FP) ghosts so that the confinement cannot be seen in perturbation theory. 
The pioneering work has been already done by Kugo \cite{Kugo}, but our new insight is that 
in order to formulate the issue of the bound states it is more convenient to make use of 
the superfield formalism in that all bound states can be summarized into a single superfield, 
thereby making it possible to derive an effective Lagrangian and field equations for asymptotic fields.
 
Before making our idea explicit, let us recall some elementary facts relevant to the later argument. 
The Hilbert space ${\cal{V}}$ defined by the Lagrangian (\ref{YM-Lag1}) has an indefinite metric,
so the S-matrix invariant and positive semi-definite physical Hilbert subspace ${\cal{V}}_{phys}$
is defined by the physical state condition \cite{Kugo-Ojima}:
\begin{eqnarray}
{\cal{V}}_{phys} = \{ |{\rm{phys}} \rangle \in {\cal{V}} \, | \, Q_B | {\rm{phys}} \rangle 
= \bar Q_B | {\rm{phys}} \rangle = 0 \},
\label{Phy-st}  
\end{eqnarray}
where let us recall that $Q_B$ and $\bar Q_B$ are respectively the BRST charge and anti-BRST charge.
To each $| {\rm{phys}}, n \rangle \in {\cal{V}}$ with ghost number $n$, there always exists 
a FP-conjugate state $| {\rm{phys}}, - n \rangle \in {\cal{V}}$ with opposite ghost number $-n$
such that their inner product is non-vanishing
\begin{eqnarray}
\langle {\rm{phys}}, -n | {\rm{phys}}, n \rangle \neq 0. 
\label{Inn-pro}  
\end{eqnarray}

We are now ready to focus on confinement of the gluon. The BRST transformation of the gluon field
$A_\mu^a$ takes the form:
\begin{eqnarray}
\delta_B A_\mu^a \equiv [ i Q_B, A_\mu^a ] = D_\mu C^a
= \partial_\mu C^a + g ( A_\mu \times C )^a \equiv \Gamma_\mu^a. 
\label{BRST-gluon}  
\end{eqnarray}
The key assumption is that the composite operator $( A_\mu \times C )^a$ develops a bound state of the vector field,
which is denoted as $\Gamma_\mu^a$, by the strong interaction. Let us note that the operators $A_\mu^a$
and $\Gamma_\mu^a$ have the same quantum properties such as mass and spin, but they
are not always massless. In fact, we will show that both the gluon and the bound states
become massive in the confinement phase.

Furthermore, let us consider the anti-BRST transformation of the gluon field
\begin{eqnarray}
\bar \delta_B A_\mu^a \equiv [ i \bar Q_B, A_\mu^a ] = D_\mu \bar C^a
\equiv \bar \Gamma_\mu^a, 
\label{anti-BRST-gluon}  
\end{eqnarray}
and its BRST transformation
\begin{eqnarray}
\delta_B \bar \delta_B A_\mu^a \equiv \{ i Q_B, [ i \bar Q_B, A_\mu^a ] \} 
= i D_\mu B^a + g ( D_\mu C \times \bar C )^a
\equiv i B_\mu^a.
\label{BRST-gluon2}  
\end{eqnarray}
Each pair of the two sets of operators, $( A_\mu, B_\mu )$ and $( C, \bar C )$ is in relation with
the FP-conjugate operator. We also assume the existence of bound states $\bar \Gamma_\mu^a$
and $B_\mu^a$.  Then, the asymptotic fields, $a_\mu, \gamma_\mu, \bar \gamma_\mu$ and
$\beta_\mu$,  which respectively correspond to $A_\mu, \Gamma_\mu, \bar \Gamma_\mu$ and
$B_\mu$, constitute a BRST quartet, and thus all members of this quartet appear only
in zero-norm combinations in the physical state ${\cal{V}}_{phys}$ and consequenly the gluon cannot be observed. 
This phenomenon is regarded as confinement of the gluon \cite{Kugo}.

It is valuable to notice that these bound states in addition to the gluon field fit into the gauge 
superfield neatly:
\begin{eqnarray}
A_\mu (z) &=&  A_\mu (x) + \theta D_\mu C (x) + \bar \theta D_\mu \bar C (x) 
+ \bar \theta \theta ( i D_\mu B (x) + g D_\mu C (x) \times \bar C (x) )
\nonumber\\
&=& A_\mu (x) + \theta \Gamma_\mu (x) + \bar \theta \bar \Gamma_\mu (x) 
+ \bar \theta \theta i B_\mu (x).
\label{BS-superfield}  
\end{eqnarray}
Furthermore, let us express the superfield for the corresponding asymptotic fields as
\begin{eqnarray}
a_\mu (z) = a_\mu (x) + \theta \gamma_\mu (x) + \bar \theta \bar \gamma_\mu (x) 
+ \bar \theta \theta i \beta_\mu (x).
\label{AS-BS-superfield}  
\end{eqnarray}
This equation will be utilized in the next section in constructing an effective Lagrangian for 
the asymptotic fields. 

In a perfectly similar manner, we can consider confinement of quarks. In this case, 
let us take account of the following BRST and anti-BRST transformations for the quark field $\Psi$,
and assume the existence of the bound states $\Sigma, \bar \Sigma$ and $\Omega$:\footnote{We
take the notation such that fermion fields commute with $\theta, \bar \theta$ and the FP ghosts.}
\begin{eqnarray}
\delta_B \Psi &\equiv& [ i Q_B, \Psi  ] = i g C \Psi \equiv \Sigma,
\nonumber\\ 
\bar \delta_B \Psi &\equiv& [ i \bar Q_B, \Psi ] =  i g \bar C \Psi 
\equiv \bar \Sigma,
\nonumber\\ 
\delta_B \bar \delta_B \Psi &\equiv& \{ i Q_B, [ i \bar Q_B, \Psi ] \} 
= - g ( B - g \bar C C ) \Psi \equiv i \Omega.
\label{BRST-quark}  
\end{eqnarray}
Actually, the existence of ghost condensation 
\begin{eqnarray}
\langle 0 | \bar C C | 0 \rangle \neq 0,
\label{Ghost-cond}  
\end{eqnarray}
is demonstrated by calculating the operator product expansion in Ref. \cite{Kondo}.
As before, these bound states and their asymptotic fields are put together into
the matter superfield:
\begin{eqnarray}
\Psi (z) &=&  \Psi (x) + \theta \Sigma (x) + \bar \theta \bar \Sigma (x) 
+ \bar \theta \theta i \Omega (x),
\nonumber\\
\psi (z) &=& \psi (x) + \theta \sigma (x) + \bar \theta \bar \sigma (x) 
+ \bar \theta \theta i \omega (x).
\label{quark-superfield}  
\end{eqnarray}
The latter equation will also be used in establishing an effective Lagrangian for the asymptotic 
fields in the next section.

\section{Effective Lagrangian for asymptotic fields}

In this section, using the superfields constructed thus far, let us make an effective Lagrangian 
for the asymptotic fields, and derive the field equations for them. As a result, we will see that 
the gluon and quark become massive dipole fields. 

Before doing so, let us point out two important remarks on the present mechanism. 
One of them is that if the bound states described in the previous section exist, 
the conventional gauge symmetry no longer ensures its BRST and anti-BRST invariances 
since we have modified them for the gluon and quark by hand. Of course, even after the
modification, both the modified BRST and anti-BRST transformations are nilpotent, 
so they play an important role in quantum field theory.   

The other important remark is that the classical QCD Lagrangian, which is composed of the Yang-Mills 
Lagrangian and the matter one, breaks the BRST and anti-BRST invariances so it cannot be part of
quantum Lagrangian any longer. Since it is not the gauge invariance but the BRST and anti-BRST invariances 
that we have to respect in taking account of the quantum field theory, we must reconsider 
an effective Lagrangian for the asymptotic fields, $(a_\mu, \gamma_\mu, \bar \gamma_\mu, \beta_\mu)$ 
and $(\psi, \sigma, \bar \sigma, \omega)$ and understand its physical implications within the framework 
of the superfield formalism. 

Now let us pay our attention to confinement of the gluon.\footnote{In the previous work \cite{Bonora-Pasti-Tonin},
the superfield formalism was adopted for understanding a violation of the cluster property in the gluon confinement 
\cite{Araki, Strocchi}.} We would like to make the effective Lagrangian obeying the following requirements: 
First, it must be invariant under the BRST and anti-BRST transformations. Note that this
requirement is automatically satisfied by taking the derivative $\frac{\partial^2}{\partial \theta \partial \bar \theta}$ 
of a superfield, which is one of advantages in the superfield formalism at hand. Secondly, the effective Lagrangian 
must be quadratic in the asymptotic superfields since they are free and non-interacting fields. Thirdly, it must contain 
the second-order derivative at most to avoid an additional ghost, which violates unitarity of the physical S-matrix. 
Finally, the kinetic term for the gauge superfield should be given with the Yang-Mills type, i.e., 
${\rm{Tr}} F_{MN} F^{MN}$, which is needed to make contact with QCD. 

The above requirements determine the effective Lagrangian for the asymptotic fields relevant to the gluon to be
\begin{eqnarray}
{\cal{L}}_{G} &=& \frac{\partial^2}{\partial \theta \partial \bar \theta} 
\Bigg[ \frac{i}{4} f^a_{\mu\nu} (z) f^{a\mu\nu} (z) + \frac{\zeta}{2} f^a_{\mu\theta} (z) f^{a\mu} \,_{\bar \theta} (z)
+ i \frac{m^2}{2} a^a_\mu (z) a^{a\mu} (z) 
\nonumber\\
&+& \frac{\alpha}{2} a^a_\theta (z) a^a_{\bar \theta} (z) \Bigg],
\label{Gluon-Lag}  
\end{eqnarray}
where $\zeta, m^2$ and $\alpha$ are constants, and we have defined
\begin{eqnarray}
f_{\mu\nu} (z) &\equiv& \partial_\mu a_\nu (z) - \partial_\nu a_\mu (z), \qquad
f_{\mu\theta} (z) \equiv \partial_\mu a_\theta (z) - \partial_\theta a_\mu (z),
\nonumber\\
f_{\mu \bar \theta} (z) &\equiv& \partial_\mu a_{\bar \theta} (z) - \partial_{\bar \theta} a_\mu (z).
\label{Def-f}  
\end{eqnarray}
Here $a_\mu (z)$ is defined in (\ref{AS-BS-superfield}), and $a_\theta (z)$ and $a_{\bar \theta} (z)$ 
take the form:
\begin{eqnarray}
a_\theta (z) = c (x) - \bar \theta i b (x),   \qquad
a_{\bar \theta} (z) = \bar c (x) + \theta i b (x),
\label{2-a}  
\end{eqnarray}
where we have used $\bar b (x) = - b (x)$ holding for the asymptotic field. Incidentally, 
from (\ref{2-a}) it is easy to verify that $f_{\theta\theta} (z) = f_{\theta\bar\theta} (z) 
= f_{\bar\theta\bar\theta} (z) = 0$. 

By virtue of these relations, the effective Lagrangian (\ref{Gluon-Lag}) can be cast to the form 
in terms of the component fields:
\begin{eqnarray}
{\cal{L}}_{G} &=& - f^a_{\mu\nu} \partial^\mu \beta^{a\nu} 
+ \frac{i}{2} ( \partial_\mu \gamma^a_\nu - \partial_\nu \gamma^a_\mu )
( \partial^\mu \bar \gamma^{a\nu} - \partial^\nu \bar \gamma^{a\mu} ) 
\nonumber\\
&+& \frac{\zeta}{2} ( \beta^a_\mu - \partial_\mu b^a )^2
- m^2 ( a^a_\mu \beta^{a\mu} - i \gamma^a_\mu \bar \gamma^{a\mu} )
+ \frac{\alpha}{2} ( b^a )^2.
\label{Gluon-Lag2}  
\end{eqnarray}
Actually, it is easy to show explicitly that this effective Lagrangian is invariant under both BRST and anti-BRST 
transformations, which take the form for the asymptotic fields:
\begin{eqnarray}
&{}& \delta_B a_\mu = \gamma_\mu, \qquad
\bar \delta_B a_\mu = \bar \gamma_\mu, \qquad
\delta_B \gamma_\mu = \bar \delta_B \bar \gamma_\mu = 0,
\nonumber\\
&{}& \delta_B \bar \gamma_\mu = i \beta_\mu, \qquad
\bar \delta_B \gamma_\mu = - i \beta_\mu, \qquad
\delta_B \beta_\mu = \bar \delta_B \beta_\mu = 0, 
\nonumber\\
&{}& \delta_B \bar c = i b, \qquad 
\bar \delta_B c = - i b, \qquad 
\delta_B c = \bar \delta_B \bar c = 0,
\nonumber\\
&{}& \delta_B b = \bar \delta_B b = 0.
\label{Gluon-BRST}  
\end{eqnarray}
Since ${\cal{L}}_{G}$ is quadratic in $\beta_{\mu\nu}$, after shifting $\beta_\mu \rightarrow \beta_\mu + \partial_\mu b$,
one can perform the path integration over $\beta_{\mu\nu}$ whose result reads
\begin{eqnarray}
{\cal{L}}_{G} &=& - \frac{1}{2 \zeta} [ ( \partial^\mu f^a_{\mu\nu} )^2 + m^4 ( a^a_\mu )^2
+ m^2 ( f^a_{\mu\nu} )^2 ] 
+ \frac{i}{2} ( \partial_\mu \gamma^a_\nu - \partial_\nu \gamma^a_\mu )
( \partial^\mu \bar \gamma^{a\nu} - \partial^\nu \bar \gamma^{a\mu} )
\nonumber\\
&-& m^2 ( a^a_\mu \partial^\mu b^a - i \gamma^a_\mu \bar \gamma^{a\mu} )
+ \frac{\alpha}{2} ( b^a )^2.
\label{Gluon-Lag3}  
\end{eqnarray}

From the effective Lagrangian (\ref{Gluon-Lag3}), it is straightforward to derive the field equations.
First of all, taking variation with respect to $b$ leads to 
\begin{eqnarray}
m^2 \partial_\mu a^\mu + \alpha b = 0.
\label{b-eq}  
\end{eqnarray}
Next, we see that the equations of motion for $\gamma_\mu$ and $\bar \gamma_\mu$ take the
similar form:
\begin{eqnarray}
( \Box - m^2 ) \gamma_\mu &=& 0, \qquad \partial_\mu \gamma^\mu = 0,
\nonumber\\
( \Box - m^2 ) \bar \gamma_\mu &=& 0, \qquad \partial_\mu \bar \gamma^\mu = 0.
\label{gamma-eq}  
\end{eqnarray}
These field equations are nothing but two massive Klein-Gordon equations with the Lorenz condition.
Moreover, with the help of Eq. (\ref{b-eq}), the field equation for $a_\mu$ becomes
\begin{eqnarray}
( \Box - m^2 )^2 a_\mu + \frac{\alpha}{m^2} \Box \partial_\mu b + (\zeta m^2 - 2 \alpha)
\partial_\mu b = 0.
\label{a-eq}  
\end{eqnarray}
Taking the divergence of this equation and using Eq. (\ref{b-eq}), we have  
\begin{eqnarray}
\Box b - \frac{\alpha}{\zeta} b = 0.
\label{Box-b}  
\end{eqnarray}
Substituting this result into Eq. (\ref{a-eq}), we arrive at
\begin{eqnarray}
( \Box - m^2 )^2 a_\mu + \frac{1}{\zeta m^2} (\alpha - \zeta m^2)^2 \partial_\mu b = 0.
\label{a-eq2}  
\end{eqnarray}
 
Here we wish to understand the physical meaning of Eq. (\ref{a-eq2}). To do that, it is
useful to compare Eq. (\ref{a-eq2}) with the field equation for the gauge field in the Yang-Mills theory,
for which we have the Klein-Gordon equation, $\Box A_\mu = 0$ in the Feynman gauge 
($\alpha = 1$).\footnote{In the Landau gauge ($\alpha = 0$), we obtain the dipole equation, 
$\Box^2 A_\mu = 0$, but this case can be treated in a similar way to the case of 
the Feynman gauge \cite{Kugo-Ojima}.} In our ``Feynman gauge'' $\alpha = \zeta m^2$,
Eq. (\ref{a-eq2}) reduces to\footnote{With this gauge choice, Eq. (\ref{Box-b}) becomes
$( \Box - m^2 ) b = 0$.}  
\begin{eqnarray}
( \Box - m^2 )^2 a_\mu = 0,
\label{a-eq3}  
\end{eqnarray}
which implies that the massless gluon has changed to the massive gluon with mass $m$ from the contribution
of the mass term, $i \frac{m^2}{2} a^a_\mu (z) a^{a\mu} (z)$ in the Lagrangian (\ref{Gluon-Lag}). 
In the presence of the gauge symmetry, such a mass term is not allowed to exist in the Lagrangian,
but owing to its absence, we have the mass term which is invariant under the BRST and anti-BRST
transformations. This situation should be compared to the usual case where the mass term is
generated dynamically through quantum corrections \cite{Chaichian, Chaichian-PoS, Kondo2}.  
It is stressed in Ref. \cite{Chaichian} that in the confinement phase color confinement is
realized and the gluon becomes massive whereas in the deconfinement phase color confinement is
not realized and the gluon remains massless. It is remarkable that our theory at hand, exhibits 
the same phenomenon. To put it differently, in the confinement phase, the mass gap naturally
appears for the gluon \cite{Witten}.

Another interesting point of Eq. (\ref{a-eq3}) is that although the gluon field has become
massive, the equation is not the conventional mass Klein-Gordon field but the massive
dipole field. It is known that the massive dipole field is constructed out of two massive
simple pole fields as follows: The single massive dipole field $a_\mu$ can be expressed by introducing an additional 
massive field $k_\mu$ like
\begin{eqnarray}
( \Box - m^2 ) a_\mu = k_\mu, \qquad
( \Box - m^2 ) k_\mu = 0.
\label{Dipole}  
\end{eqnarray}
Then, the massive dipole field $a_\mu$ is described as
\begin{eqnarray}
a_\mu = \hat a_\mu + \frac{1}{2 m^2} ( x^\nu \partial_\nu + c_0 ) k_\mu,
\label{Dipole2}  
\end{eqnarray}
where $( \Box - m^2 ) \hat a_\mu =0$ and $c_0$ is an arbitrary constant.
Changing the value of $c_0$ is equivalent to adding $\hat a_\mu$ to $a_\mu$,
so we can set $c_0 = 1$ without loss of generality. 

In this way, we find that the massive dipole field $a_\mu$ can be described in terms of a pair of massive
simple pole fields, $k_\mu$ and $\hat a_\mu$. In other words, in order that the massless gluon 
is confined in a hadron, not only it must acquire a new degree of freedom to become massive
but also it needs to have another massive field to constitute a bound state. It is of interest
to recall that in particular the latter feature emerges in quadratic gravity when massive ghost makes a bound state,
thereby nulltifying the problem of unitarity violation of the physical S-matrix \cite{Oda-MG}.      

Finally, it is worthwhile to notice that all members of the BRST quartet, $(a_\mu, \gamma_\mu, 
\bar \gamma_\mu, \beta_\mu)$ share the same magnitude of mass as expected from the BRST
and anti-BRST invariances. In fact, the field equation for $\beta_\mu$ provides
\begin{eqnarray}
\beta_\mu = \frac{1}{\zeta} [ - ( \Box - m^2 ) a_\mu + \partial_\mu \partial_\nu a^\nu ].
\label{Beta-eq}  
\end{eqnarray}
Using Eqs. (\ref{b-eq}), (\ref{a-eq3}), and $( \Box - m^2 ) b = 0$, we can obtain 
\begin{eqnarray}
( \Box - m^2 ) \beta_\mu =0.
\label{Beta-eq2}  
\end{eqnarray}
Although the magnitude of mass of these fields is in general modified
by radiative corrections, every member of the same quartet receives the same mass renormalization
and always takes the same magnitude of mass owing to the BRST and anti-BRST invariances 
as long as these invariances are not broken.  

Next, we will move on to the issue of confinement of the quark. At this stage, we meet another 
disadvantage of the superfield formalism; there is no way of incorporating the Lagrangian 
for matter fields in six-dimensional superspace. This situation is similar to that of massive ghost 
in quadratic gravity \cite{Oda-MG}, so we shall follow a more pragmatic approach as in quadratic gravity.

We will therefore adopt the slightly modified requirements:
First, it must be invariant under the BRST and anti-BRST transformations which is achieved by taking 
the derivative $\frac{\partial^2}{\partial \theta \partial \bar \theta}$ of a superfield. 
Secondly, the effective Lagrangian must be quadratic in the asymptotic superfields and must be
a bilinear form of $\partial_\theta \Psi (z)$ and $\partial_{\bar \theta} \Psi (z)$  
since they are free and non-interacting fields. Thirdly, it must contain 
the second-order derivative at most to avoid an additional ghost. Finally, the kinetic term 
for the spinor superfield should be given with the four-dimensional Dirac type, 
i.e., $\tilde \Psi (z) ( i \Slash{\partial} - m ) \Psi (z)$, which is needed to make
contact with QCD. This final requirement is a little unnatural since we introduce 
the Dirac-type of Lagrangian not in six-dimensional superspace but in four-dimensional
space-time.    
 
Following these requirements, let us make an effective Lagrangian for asymptotic fields relevant to 
the quark confinement and derive the equations of motion from it. Our argument proceeds 
along the same line as for the gluon confinement with some minor technical complications 
associated with spinor fields, and leads to a similar result to the gluon confinement.    

From the requirements, we consider the Dirac Lagrangian made from the spinor superfield 
in Eq. (\ref{quark-superfield}):
\begin{eqnarray}
{\cal{L}}_{D} &=& \tilde \Psi (z) ( i \gamma^\mu D_\mu - m ) \Psi (z)
\nonumber\\
&=& \tilde \Psi (z) [ i \gamma^\mu ( \partial_\mu + A_\mu ) - m ] \Psi (z),
\label{Dirac-Lag}  
\end{eqnarray}
where $\tilde \Psi \equiv \Psi^\dagger \gamma^0$ is the Dirac conjugate and the Dirac matrices 
$\gamma^\mu$ satisfy the Clifford algebra, $\{ \gamma^\mu, \gamma^\nu \} = - 2 \eta^{\mu\nu}$. 
For the asymptotic fields $\psi (z)$ in Eq. (\ref{quark-superfield}), the Lagrangian (\ref{Dirac-Lag}) reduces to
\begin{eqnarray}
{\cal{L}}_{D} &=& \tilde \psi (z) ( i \Slash{\partial} - m ) \psi (z)
\nonumber\\
&=& \bar \theta \theta \Bigg[ \tilde \psi (x) ( i \Slash{\partial} - m ) i \omega (x)  
+ \tilde \sigma (x) ( i \Slash{\partial} - m ) \bar \sigma (x) 
\nonumber\\
&-& \tilde{\bar \sigma} ( i \Slash{\partial} - m ) \sigma (x) 
+ i \tilde \omega (x) ( i \Slash{\partial} - m ) \psi (x) \Bigg] + \dots,
\label{AS-Dirac-Lag}  
\end{eqnarray}
where $\Slash{\partial} \equiv \gamma^\mu \partial_\mu$ and $\dots$ indicates
the terms independent of $\bar \theta \theta$. The other admissible candidate as the Lagrangian
for the spinor superfield is
\begin{eqnarray}
{\cal{L}}_{\omega} &=& \partial_\theta \tilde \psi (z) \partial_{\bar \theta} \psi (z)
= \bar \theta \theta \tilde \omega (x) \omega (x) + \dots.
\label{Sp-cand}  
\end{eqnarray}

Thus, we have the BRST and anti-BRST invariant Lagrangian by picking up
component field associated with $\bar \theta \theta$ as
\begin{eqnarray}
{\cal{L}}_{Q} &=& \frac{\partial^2}{\partial \theta \partial \bar \theta} 
\left( {\cal{L}}_{D} + {\cal{L}}_{\omega} \right)
\nonumber\\
&=& \tilde \psi (x) ( i \Slash{\partial} - m ) i \omega (x)  
+ \tilde \sigma (x) ( i \Slash{\partial} - m ) \bar \sigma (x) 
- \tilde{\bar \sigma} ( i \Slash{\partial} - m ) \sigma (x) 
\nonumber\\
&+& i \tilde \omega (x) ( i \Slash{\partial} - m ) \psi (x) + \tilde \omega (x) \omega (x).
\label{Quark-Lag}  
\end{eqnarray}
Actually, this Lagrangian is invariant under the following BRST and anti-BRST transformations:
\begin{eqnarray}
\delta_B \psi &=& \sigma, \qquad
\bar \delta_B \psi = \bar \sigma, \qquad
\delta_B \sigma = \bar \delta_B \bar \sigma = 0,
\nonumber\\
\delta_B \bar \sigma &=& i \omega, \qquad
\bar \delta_B \sigma = - i \omega, \qquad
\delta_B \omega = \bar \delta_B \omega = 0.
\label{Quark-BRST}  
\end{eqnarray}
In particular, note that $\delta_B \tilde{\bar \sigma} = i \tilde \omega$, which can be read off from 
the equation:
\begin{eqnarray}
\tilde \psi (z) &=& \tilde \psi (x) + \theta \tilde \sigma (x) + \bar \theta \tilde{\bar \sigma}
+ \bar \theta \theta i \tilde \omega (x)
\nonumber\\
&=& \tilde \psi (x) + \theta \delta_B \tilde \psi (x) + \bar \theta \bar \delta_B \tilde \psi (x)
+ \bar \theta \theta \delta_B \bar \delta_B \tilde \psi (x).
\label{Quark-BRST2}  
\end{eqnarray}

From the effective Lagrangian (\ref{Quark-Lag}), it is straightforward to derive the whole equation of motion:
\begin{eqnarray}
( i \Slash{\partial} - m )^2 \psi (x) &=& 0, \qquad
( i \Slash{\partial} - m ) \omega (x) = 0,
\nonumber\\
( i \Slash{\partial} - m ) \sigma (x) &=& 0, \qquad 
( i \Slash{\partial} - m ) \bar \sigma (x) = 0.
\label{Quark-Eq}  
\end{eqnarray}
Note that the quark field $\psi$ satisfies the massive dipole spinor equation whereas the other fields
do the conventional massive Dirac equation. As in the massive dipole field $a_\mu$ 
in Eq. (\ref{a-eq3}),  the massive dipole spinor field $\psi$ can be expressed by a pair of
massive Dirac fields $\chi$ and $\hat \psi$ as follows: The field equation for $\psi$ can be
decomposed into two equations:
\begin{eqnarray}
( i \Slash{\partial} - m ) \psi = \chi, \qquad
( i \Slash{\partial} - m ) \chi = 0.
\label{Dipole-sp}  
\end{eqnarray}
Then, it is easy to check that the massive dipole spinor field $\psi$ is described as
\begin{eqnarray}
\psi = \hat \psi + \frac{1}{m} ( x^\mu \partial_\mu + c_1 ) \chi,
\label{Dipole2-sp2}  
\end{eqnarray}
where $( i \Slash{\partial} - m ) \hat \psi =0$ and $c_1$ is an arbitrary constant.
Changing the value of $c_1$ amounts to adding $\hat \psi$ to $\psi$,
so we can take $c_1 = 1$ without loss of generality as before.

To sum up, all members of the quartet have the same mass $m$, and satisfy the massive
Dirac equation except for the quark field $\psi (x)$, which obeys the massive
dipole spinor equation. It is of interest that all fundamental fields of the gluon and
the quark in QCD as well as massive ghost in quadratic gravity take the form of the dipole
field in the confinement phase. In particular, we can provide the following interesting picture
of the quark confinement: Since a massive dipole spinor is in general composed of two massive Dirac spinors, 
which might be identified with a quark and anti-quark, a pair of the quark and anti-quark is confined,
thereby generating a bound state called a meson.

\section{Conclusions}

In this article, inspired by our recent work of quadratic gravity \cite{Oda-MG}, we have reconsidered 
color confinement in QCD in the superfield formalism by Bonora and Tonin \cite{Bonora-Tonin}. The present 
article simply insists on two points: The one point is that the gluon, which is originally massless owing to the $SU(3)$
local gauge symmetry, must become massive in the confinement phase. Namely, when the gluon is confined by
the strong interaction, the gluon natually develops mass gap in quantum field theory without recourse to some dynamical 
mass generation mechanism. The other point is that the gluon and the quark as well as the massive ghost must obey 
the massive dipole field equation in the confinement phase.  

Let us dwell on the above two points. The former point, i.e., the emergence of the massive gluon in the confinement phase,
seems to be physically reasonable. From the viewpoint of special relativity, it is very difficult, even if it is not impossible,
to confine a massless particle, which moves at the velocity of light, into a small region such as inside a hadron 
even by the strong interaction. From the viewpoint of quantum field theory, it is also difficult to do so since confining 
the massless particle within the small region gives the huge kinetic energy to the massless particle owing to 
the Heisenberg's uncertainty principle. In other words, all the particles must somehow acquire mass if they are
confined by some interaction. Our present formalism clearly demonstrates this fact.

The latter point, i.e., the emergence of not the simple pole field but the dipole field, is also physically interesting.
As shown in the article, a single massive dipole field is composed of a pair of massive simple pole fields.     
Thus, the emergence of the dipole field implies that in order to a bound state, at least a pair of the simple pole fields is needed.
In particular, this situation nicely matches the picture such that a pair of quark and anti-quark makes a bound state, which is
a meson. An interesting question is how to describe baryons in terms of the present formalism. If our interpretation is 
correct, is it necessary to consider tripole fields to describe the baryons?

One of the remaining important problems is to show that in several channels such as $g ( A_\mu \times C)$, there appear 
bound states. In this respect, it is encouraging for us to know the fact that many of bound states such as the ghost condensation 
and the gluon condensation are generated in QCD. Another important problem is to understand confinement of massive ghost in
quadratic gravity \cite{Oda-MG} within the framework of the manifestly covariant canonical formalism \cite{Oda-Can}.
In order to do so, we wish to borrow some useful techniques and knowledge developed for color confinement in
QCD and utilize the approach found in recent works \cite{Oda0, Oda1}.


\end{document}